\documentclass[apj]{emulateapj}
\usepackage{amsmath}
\usepackage{txfonts}
\eqsecnum

\renewcommand\email\texttt

\newcommand\coords{{08:51:30\, +63:07:48}}

\def\spose#1{\hbox to 0pt{#1\hss}}
\def\lta{\mathrel{\spose{\lower 3pt\hbox{$\sim$}}
    \raise 2.0pt\hbox{$<$}}}
\def\gta{\mathrel{\spose{\lower 3pt\hbox{$\sim$}}
    \raise 2.0pt\hbox{$>$}}}

\begin{document} 

\slugcomment{\sc submitted to \it Astrophysical Journal Letters}
\shorttitle{\sc Ursa Major II Dwarf} 
\shortauthors{Zucker et al.}

\title{A Curious New Milky Way Satellite in Ursa Major\footnotemark[0]}

\author{D.\ B.\ Zucker\altaffilmark{1}, 
V.\ Belokurov\altaffilmark{1},
N.\ W.\ Evans\altaffilmark{1}, 
J.\ T.\ Kleyna\altaffilmark{2},
M.\ J.\ Irwin\altaffilmark{1},
M.\ I.\ Wilkinson\altaffilmark{1},
M.\ Fellhauer\altaffilmark{1},
D.\ M.\ Bramich\altaffilmark{1},
G.\ Gilmore\altaffilmark{1},
H.\ J.\ Newberg\altaffilmark{3},
B.\ Yanny\altaffilmark{4},
J.\ A.\ Smith\altaffilmark{5,6},
P.\ C.\ Hewett\altaffilmark{1},
E.\ F.\ Bell\altaffilmark{7},
H.-W.\ Rix\altaffilmark{7},
O.\ Y.\ Gnedin\altaffilmark{8},
S.\ Vidrih\altaffilmark{1}, 
R.\ F.\ G. Wyse\altaffilmark{9},
B.\ Willman\altaffilmark{10},
E.\ K.\ Grebel\altaffilmark{11},
D.\ P.\ Schneider\altaffilmark{12},
T.\ C.\ Beers\altaffilmark{13},
A.\ Y.\ Kniazev\altaffilmark{7,14}
J.\ C.\ Barentine\altaffilmark{15},
H.\ Brewington\altaffilmark{15},
J.\ Brinkmann\altaffilmark{15},
M.\ Harvanek\altaffilmark{15},
S.\ J.\ Kleinman\altaffilmark{16},
J.\ Krzesinski\altaffilmark{15,17},
D.\ Long\altaffilmark{15},
A.\ Nitta\altaffilmark{18},
S.\ A.\ Snedden\altaffilmark{15}
}

\altaffiltext{1}{Institute of Astronomy, University of Cambridge,
Madingley Road, Cambridge CB3 0HA, UK;\email{zucker,vasily,nwe@ast.cam.ac.uk}}
\altaffiltext{2}{Institute for Astronomy, University of Hawaii, 2680
  Woodlawn Drive, Honolulu, HI 96822}
\altaffiltext{3}{Rensselaer Polytechnic Institute, Troy, NY 12180}
\altaffiltext{4}{Fermi National Accelerator Laboratory, P.O. Box 500, Batavia, IL 60510}
\altaffiltext{5}{Los Alamos National Laboratory, ISR-4, MS D448, Los Alamos, NM 87545}
\altaffiltext{6}{Department of Physics and Astronomy, Austin Peay State University, P.O. Box 4608, Clarksville, TN 37040} 
\altaffiltext{7}{Max Planck Institute for Astronomy, K\"{o}nigstuhl 17, 69117 Heidelberg, Germany}
\altaffiltext{8}{Department of Astronomy, Ohio State University, 140 West 18th Avenue, Columbus, OH 43210}
\altaffiltext{9}{The Johns Hopkins University, 3701 San Martin Drive, Baltimore, MD 21218}
\altaffiltext{10}{Center for Cosmology and Particle Physics, Department of Physics, New York University, 4 Washington Place, New York, NY 10003}
\altaffiltext{11}{Astronomical Institute of the University of Basel, Department of Physics and Astronomy, Venusstrasse 7, CH-4102 Binningen, Switzerland}
\altaffiltext{12}{Department of Astronomy and Astrophysics, Pennsylvania State University, 525 Davey Laboratory, University Park, PA 16802}
\altaffiltext{13}{Department of Physics and Astronomy, 
Michigan State University, East Lansing, MI 48824}
\altaffiltext{14}{South African Astronomical Observatory, PO Box 9, Observatory 7935, Cape Town, South Africa}
\altaffiltext{15}{Apache Point Observatory, P.O. Box 59, Sunspot, NM 88349}
\altaffiltext{16}{Subaru Telescope, 650 N. A'ohoku Place, Hilo, HI 96720}
\altaffiltext{17}{Mt.\ Suhora Observatory, Cracow Pedagogical University, ul.\ Podchorazych 2, 30-084 Cracow, Poland}
\altaffiltext{18}{Gemini Observatory, 670 N. A'ohoku Place, Hilo, HI 96720}
\footnotetext[0]{Based in part on data collected at Subaru Telescope, which is operated by the National Astronomical Observatory of Japan.}

\begin{abstract}
In this Letter, we study a localized stellar overdensity in the
constellation of Ursa Major, first identified in Sloan Digital Sky
Survey (SDSS) data and subsequently followed up with Subaru imaging.
Its color-magnitude diagram (CMD) shows a well-defined sub-giant branch,
main sequence and turn-off, from which we estimate a distance of $\sim
30$ kpc and a projected size of $\sim 250 \times 125$ pc. The CMD suggests a composite population with some range in metallicity and/or age. Based on its extent
and stellar population, we argue that this is a previously unknown
satellite galaxy of the Milky Way, hereby named Ursa Major II (UMa II) after its
constellation.  Using SDSS data, we find an
absolute magnitude of $M_V \sim -3.8$, which would make it the
faintest known satellite galaxy. UMa II's isophotes are irregular and
distorted with evidence for multiple concentrations; this suggests
that the satellite is in the process of disruption.
\end{abstract}


\section{Introduction}

Numerical simulations in the hierarchical cold dark matter paradigm
of galaxy formation generally predict 1 to 2 orders of
magnitude more satellite halos in the present day Local Group than the
number of dwarf galaxies thus far observed
\citep[e.g.,][]{Mo99,Kl99,Be02}. Numerous solutions have been proposed
for this ``missing satellite'' problem. For example, star formation
may be inhibited in low-mass systems \citep[e.g.,][]{Bu01,So02}, or
the known satellites may represent a higher mass regime of the
satellite initial mass function~\citep[e.g.,][]{Str02,Kr04}.

However, it has become increasingly clear over the last two years that
the census of Local Group satellites is seriously incomplete. Data
from the Sloan Digital Sky Survey \citep[SDSS;][]{Yo00} have revealed
five new nearby dwarf spheroidals (dSphs) in quick succession:
Andromeda IX~\citep{Zu04}, Ursa Major~\citep{Wi05a}, Andromeda
X~\citep{Zu06b}, Canes Venatici~\citep{Zu06a} and
Bo{\"o}tes~\citep{Be06b}. All five galaxies were detected as
stellar overdensities. The purpose of this Letter is to study another
prominent stellar overdensity in SDSS Data Release 4 \citep{Am06}.
\citet{Gr06} independently called attention to it, stating that it may be a ``new globular cluster or dwarf spheroidal''.
Here we provide evidence from SDSS and subsequent deeper Subaru
imaging for its interpretation as a dwarf spheroidal galaxy, the
thirteenth around the Milky Way, with the proposed name Ursa Major II (UMa II).

\begin{figure}[t]
\begin{center}
\epsscale{1.0}
\plotone{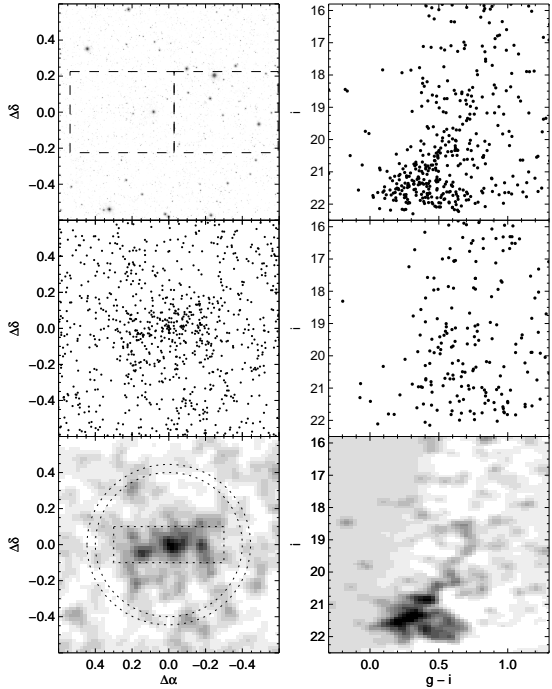}
\caption{The UMa II Dwarf as seen by SDSS: {\it Upper left:} Combined
SDSS $g,r,i$ images of a $1.2^{\circ} \times 1.2^{\circ}$ field
centered on the overdensity (J2000 08:51:30 +63:07:48). $\Delta
\alpha$ and $\Delta \delta$ are the relative offsets in right
ascension and declination, measured in degrees of arc. The dashed
lines indicate the two pointings observed with Subaru (see \S 2).
{\it Middle left:} The spatial distribution of all blue objects ($g -
i < 0.5$) classified as stars in the same area. {\it Lower left:}
Binned spatial density of all blue stellar objects, together with a
dotted box that covers most of the object and a dotted annulus used to
define the background. {\it Upper right:} CMD of all stellar objects
within the dotted box; note the clear main sequence turn-off and
subgiant branch, along with hints of horizontal and red giant
branches, even without removal of field contamination. {\it Middle right:} Control CMD of field stars from the dotted annulus. {\it
Lower right:} A color-magnitude density plot (Hess diagram), showing
the CMD of the box minus the control CMD, normalized to the number of
stars in each CMD. All photometric data were corrected for Galactic
foreground extinction using \citet{schl98}.
\label{fig:uma_disc}}
\end{center}
\vskip -1.1cm
\end{figure}
\begin{figure}
\begin{center}
\epsscale{1.0}
\plotone{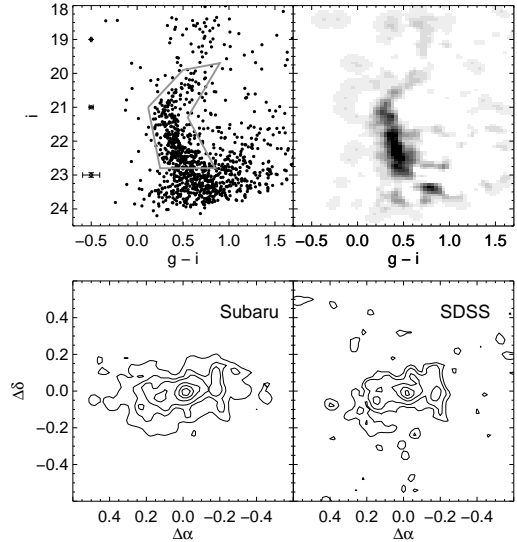}
\caption{The UMa II Dwarf as seen by Subaru: {\it Upper left:} CMD of
the central region of UMa II (see dashed box in the upper right panel
of Figure~\ref{fig:uma_disc}), constructed with Subaru $g,i$ data. The
solid gray line graphically indicates the color-magnitude selection
criteria used to construct the contour plots in the lower panels. The error bars on the left show the typical photometric errors at the $i-$band magnitude indicated. {\it
Upper right:} A color-magnitude density plot (Hess diagram), showing
the CMD of the box minus a control-field CMD, normalized to the number
of stars in each CMD. {\it Lower left:} Isodensity contours of the
stars selected from the Subaru data by the gray box in the upper left
panel. The plotted contour levels are 1, 2, 3, 5, 7 and 9$\sigma$
above the background level. $\Delta \alpha$ and $\Delta \delta$ are
measured in degrees of arc.  {\it Lower right:} Isodensity contours
using SDSS data for comparison, with levels of 2, 3, 5, 7 and
9$\sigma$ above the background plotted. Note that the three blobs
appear in both panels.}
\label{fig:uma_disc2}
\end{center}
\vskip -1.2cm
\end{figure}
\begin{figure}[t]
\begin{center}
\epsscale{1.0}
\plotone{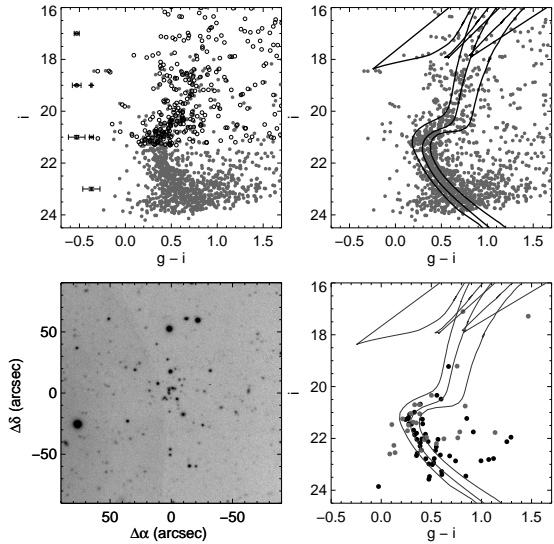}
\caption{ {\it Upper left:} Composite CMD of the central region of UMa
II (dashed box in the upper right panel of Figure~\ref{fig:uma_disc}),
with photometry of both SDSS and Subaru stars plotted (black circles
and gray dots, respectively). Duplicate detections (i.e., detections
of the same star in both sets of photometric data) have {\em not} been
removed. The error bars on the left show the typical photometric
errors for each dataset at the $i-$band magnitude indicated. {\it
Upper right:} The same composite CMD, with all stars shown as gray
dots, and Padova isochrones \citep{Gi04} overplotted for (left to
right) [Fe/H]$= -2.3$/12 Gyr, [Fe/H]$= -1.3$/12 Gyr and [Fe/H]$=
-0.7$/10 Gyr, shifted to a distance modulus of 17.5.  {\it Lower
left:} Subaru $g-$band image of the apparent central cluster of UMa
II. The image spans $3\arcmin \times 3\arcmin$. The curved shadow to
the left is scattered light from a nearby bright star.  {\it Lower
right:} Composite CMD of the central cluster region shown in the lower
left panel, with SDSS and Subaru photometry plotted as gray and black
dots, respectively. The three isochrones from the upper right panel
are also overplotted; the middle isochrone ([Fe/H]$= -1.3$/12 Gyr)
appears to be a reasonably good fit to the data, although even in this
small region the main sequence is broader than might be expected from
simple photometric errors (see upper left panel).}
\label{fig:uma_cmdscluster}
\end{center}
\vskip -0.5cm
\end{figure}
\begin{figure}
\begin{center}
\epsscale{0.95}
\plotone{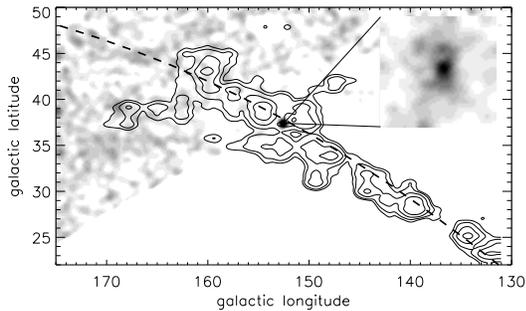}
\caption{The locations of UMa II and Complex A, together with the
great circle of the Orphan Stream. The distance estimate to the Orphan
Stream is comparable to that of UMa II, but Complex A is believed to
lie much closer.  The gray scale shows the density of SDSS stars
satisfying $g - r < 0.4$ and $20 < r < 22.5$. The inset image is a
blow-up of the area immediately around UMa II, showing its long axis
almost aligned with constant Galactic longitude. The column density
contours for Complex A are taken from \citet{Wa01}, while the great
circle of the Orphan Stream is from \citet{Be06c}.}
\label{fig:busy}
\end{center}
\vskip -1.2cm
\end{figure}

\begin{deluxetable}{lc}
\tablecaption{Properties of the Ursa Major II Dwarf \label{tbl:pars}}
\tablewidth{0pt} \tablehead{ \colhead{Parameter\tablenotemark{a}} &
{~~~ } } \startdata Coordinates (J2000) & \coords \\
Coordinates (Galactic) & $\ell = 152.5^\circ$, $b= 37.4^\circ$ \\
 Position Angle & $95^\circ$\\
 Ellipticity & $0.5$\\
 Central Extinction, A$_{\rm V}$ & $0\fm29$\\
V$_{\rm tot}$ & $14\fm3\pm0\fm5$\\
(m$-$M)$_0$ & $17\fm5\pm0\fm3$\\
 M$_{\rm tot,V}$ & $-3\fm8\pm0\fm6$\enddata
\tablenotetext{a}{Integrated magnitudes are
corrected for the Galactic foreground reddening reported by \citet{schl98}}
\label{tab:struct}
\end{deluxetable}

\section{Observations and Data Analysis}

The SDSS is an imaging and spectroscopic survey, with imaging data
taken in five photometric bands \citep[$u$, $g$, $r$, $i$ and
$z$;][]{Fu96,Gu98,Gu06,Ho01,Am06} and automatically
processed through photometric and astrometric pipelines~\citep{Lu99,St02,Sm02,Pi03,Iv04}. As part of our systematic
analysis of SDSS data around the north Galactic pole~\citep[see,
e.g.,][]{Be06a}, we identified a stellar overdensity in the
constellation Ursa Major.

Figure~\ref{fig:uma_disc} shows a set of panels derived from the SDSS
public data.  A combined $g,r,i$ grayscale image
centered on the stellar overdensity (upper left) reveals no obvious object,
However, by selecting only the objects classified by the SDSS
pipeline as blue stars ($g - i < 0.5$), a flattened stellar
overdensity is readily visible in the photometric data (middle and bottom left panels). A color-magnitude diagram (CMD) of
all stars in the central region reveals a clear main sequence turn-off
and sub-giant branch, as well as what appear to be a red clump and
sparse horizontal and red giant branches (right panels). 
The CMD bears some resemblance to those of intermediate-metallicity globular clusters,
but the
satellite's seemingly irregular and broken morphology with distinct
blobs and sub-clumps leaves its precise nature open to question. It is
not even clear that it is a single object.

Accordingly, we obtained deeper follow-up observations with the Suprime-Cam mosaic imager~\citep{Mi02} on the Subaru
telescope~\citep{Iy04}.
Data were gathered on 2006 May 26 (UT),
using two pointings to cover the stellar overdensity (upper left panel
of Figure~\ref{fig:uma_disc}).  Each pointing was observed in $g'$ and
$i'$ bands (for ease of calibration with SDSS data) in a 3-exposure
dither to cover the gaps between CCDs. Each exposure was 240s, for 12
minutes of exposure time in each band.  Unfortunately, several
exposures of the western pointing were affected by problems with
vignetting and tracking, so that only single $g'$ and $i'$ exposures
of this area were usable. The data were processed using
a general purpose pipeline
modified for Subaru reductions. Images
were debiased and trimmed, and then flatfielded and gain-corrected to
a common internal system using clipped median stacks of nightly
twilight flats. Aperture photometry from these processed images was
then bootstrap calibrated onto the SDSS photometric system.

The upper left panel of Figure~\ref{fig:uma_disc2} shows a deep CMD
derived from our Subaru imaging. A densely populated upper main
sequence and sub-giant branch are now clearly discernible, though with
only a truncated red giant branch and possible horizontal branch
because of saturation in the Subaru data 
brighter than $i \sim 18$.  
The upper right panel of Figure~\ref{fig:uma_disc2} shows a background-subtracted Hess diagram of the object.
The solid gray line that wraps around the object's main
sequence and sub-giant branch in the upper left panel is used to
select members. The density contours derived from the spatial
distribution of these stars are shown in the lower left panel. The
central parts of the object break up into three distinct clumps. These
are also visible in the density contours derived from the SDSS data,
using the same color-magnitude selection, giving additional confidence
that they are not merely data artifacts.

\section{Properties of UMa II}
\label{sec:props}

The upper left panel of Figure~\ref{fig:uma_cmdscluster} shows a
composite CMD of the central parts of the object with the bright stars
taken from SDSS and the faint ones from Subaru. The width in the upper
main sequence far exceeds the observational errors and the expected range in
foreground extinction \citep[$\Delta E(g - i) \sim 0.1$;][]{schl98} and may be caused by a number of factors. First,
there is nebulosity in the field of our $g'$ band Subaru images,
suggesting that there may be patchy reddening unresolved on the
scale of Schlegel et al.'s (1998) maps. Second, the spread could be
caused by depth along the line of sight, although the main sequence is
nearly vertical near the turn-off and thus a distance spread alone
would not reproduce its observed width.  Finally, it could be caused
by a mix of stellar populations of different metallicity and age. This
last hypothesis is illustrated in the upper right panel of
Figure~\ref{fig:uma_cmdscluster} by the overplotting of isochrones of
different metallicities and ages from \citet{Gi04}.  The stellar
population is not well-described by a single isochrone, but the data
are consistent with a single distance, and an age/metallicity
range. Judging from the isochrones, a reasonable conclusion is that
the object is of intermediate metallicity and 
at least 10 Gyrs old.
A Subaru $g'$ band image of what appears to be a central cluster is
shown in the lower left panel of Figure~\ref{fig:uma_cmdscluster}. The
image is dominated by turn-off and sub-giant stars (lower right
panel). Even in the small area of the central cluster, the main
sequence appears to be broader than that of a single
population.

Given the breadth of the main sequence and turn-off, it is difficult
to determine a precise distance to the object. From the overlaid
isochrones, we estimate a distance modulus of $(m\!-\!M)_0 \sim 17.5\pm0.3$,
corresponding to $\sim 30\pm5$ kpc. At that distance, the angular extent of the object ($\sim 0.5^\circ \times 0.25^\circ$) translates to a size of $\sim 250 \times 125$ pc. Using the same method described in
\citet{Be06b}, we estimate its absolute magnitude as $M_V
\sim -3.8\pm0.6$, a value consistent with the absence of a
significant number of giant stars.
Based on its size (which exceeds typical values for faint globular clusters), its broad CMD morphology (which argues against a single stellar population), and its extremely low surface
brightness, we conclude that this is most likely a hitherto unknown
dSph galaxy. As it is the second Milky Way dSph satellite to be
discovered in this constellation, we follow convention in naming it
Ursa Major II (UMa II).

\section{Discussion}

At $M_V \sim -3.8$, UMa II would be the faintest
dSph yet discovered. One might therefore wonder whether
UMa II could instead be a large globular cluster with 
gross tidal distortions.  In a globular cluster
undergoing tidal disruption, the transverse size of the tail does not
increase appreciably~\citep{De04}. Thus the diameter of the globular
cluster would have to be $\sim 125$ pc, larger than almost all known globulars.
In addition, the CMD does not resemble that of a single
stellar population, as in a typical globular cluster. If UMa
II were a disrupted cluster, the progenitor would likely have had 
properties more extreme than the largest Milky Way globular, $\omega$
Centauri, itself widely believed to be the nucleus of a
dSph~\citep[e.g.,][]{Ma00}.

Yet, in its physical properties UMa II does resemble Willman~1 -- a peculiar object also found with SDSS -- which may be a tidally-disrupted globular cluster~\citep{Wi05b,Wi06}. Willman~1's absolute magnitude and half-light radius, $M_{V} \sim -2.5$ and $r_{1/2} \sim 20$ pc~\citep{Wi06}, are at least a factor of $\sim 3$ fainter and smaller than the corresponding quantities for UMa II, $M_{V} \sim -3.8$ and $r_{1/2} \sim 50$ pc or $\sim 120$ pc (based on the minor axis or azimuthally averaged). 
In the $M_{V}$ vs. $r_{1/2}$ plane,
UMa II would lie between Willman~1 and the recently discovered low-luminosity Milky Way dSph satellites Ursa Major I, Bo\"{o}tes and Canes Venatici. With digital surveys like SDSS we are thus probing a new regime of ultra-low surface brightness stellar structures, where -- in the absence of kinematic data -- the distinction between globular clusters and dwarf galaxies is no longer obvious.

If the thickness of the main sequence is indicative of episodic
or extended star formation, then UMa II may once have been much more
massive and subsequently suffered disruption. The isophotes of UMa II
are even more distorted and irregular than those of the Ursa
Minor~\citep{Pa03} or Bootes dSphs~\citep{Be06b}.  In addition to the
central cluster, there appear to be two density peaks at ($\Delta
\alpha \sim 0.2^\circ, \Delta \delta \sim 0.0^\circ$) and at
($\Delta \alpha \sim -0.2^\circ, \Delta \delta \sim
-0.05^\circ$). These may perhaps be just fragments of what was once a
regular galaxy.  If UMa II were unbound, the fragments would probably
have been completely disrupted by now and would not be detected as a
significant stellar overdensity. 

But is UMa II gravitationally bound? 
We can estimate what mass-to-light ratio (M/L)
would be required for it to be bound using the criterion for tidal disruption of a cluster of particles in a circular orbit: $3 M_{\rm MW}/D_{\rm MW}^3 > M_{\rm UMa II}/R_{\rm UMa II}^3$, where $M_{\rm MW}$ and
$M_{\rm UMa II}$ are the enclosed masses of the Milky Way and UMa II,
$D_{\rm MW}$ is the distance of UMa II from the center of the Milky Way, and
$R_{\rm UMa II}$ is the radius of UMa II. Assuming $M_{\rm MW} \sim 4\times10^{11} M_{\odot}$, $D_{\rm MW} \sim 36$ kpc (from a heliocentric distance of $\sim 30$ kpc), $R_{\rm UMa II} \sim 100$ pc, and $L_{\rm UMa II} \sim 3\times10^{3} L_{\odot}$ (based on $M_{V,UMa II} \sim -3.8$ and $M_{V,\odot} \sim 4.85$), UMa II would require M/L $\sim 8$ to be marginally bound. 
The uncertainties in this estimate are substantial -- just the approximations inherent in the formula could introduce errors of a factor of $\sim 4$ -- but it does suggest that, if UMa II is gravitationally bound, its M/L may be higher than that of a typical stellar population. 
In other words, the existence of UMa II as a presumably long-lived, distinct object could imply a higher M/L than would be expected in a system without dark matter.
 
UMa II is found in a busy area of sky, as shown in
Figure~\ref{fig:busy}.  UMa II lies on the great circle of the
``Orphan Stream'', a $\sim 50^\circ$ stellar stream discovered in SDSS
data~\citep{Be06c,Gr06}. The distance to the Orphan Stream is $\sim
30$ kpc, comparable to UMa II.  The great circle of the Orphan Stream
includes a number of anomalous, young halo globular clusters,
particularly Palomar 1 and Ruprecht 106.  UMa II also lies close to
the association of HI high velocity clouds known as Complex
A~\citep[see e.g.,][]{Wa01}. Complex A has a distance bracket of 4.0
to 10.1 kpc~\citep{Wa96}. Although this is much closer than UMa II,
nonetheless they may be associated if Complex A lies on a different
orbital wrap of a mutual progenitor~\citep{Be06c}.  UMa II could thus
be a surviving fragment of a larger progenitor.

\section{Conclusions}

We have identified a new companion to the Milky Way galaxy in the
constellation Ursa Major.  Based on its size, structure and stellar
population, we argue that it is a new dwarf spheroidal galaxy and
name it UMa II. It has a distance of $\sim 30$ kpc and an absolute
magnitude of $M_V \sim -3.8$. Its color-magnitude diagram shows an
upper main sequence, turn-off and sub-giant branch, as well as hints
of red giant and horizontal branches.  UMa II has a bright central
concentration, together with two further clumps. The irregular nature
of the object suggests that it may have undergone disruption.

This is the fourth Milky Way dSph discovered by SDSS in little over a
year. Together with the earlier discoveries of Ursa Major I,
Canes Venatici and Bo{\"o}tes, this underscores how incomplete our
current census actually is.  As SDSS covers only $\sim 1/4$
of the celestial sphere, crude scaling arguments would suggest
that there are tens of missing Milky Way dSphs.  If true, this
would go some way toward resolving the missing satellite issue.

\vskip-0.5cm

\acknowledgments 
Funding for the SDSS and SDSS-II has been provided by the Alfred P.
Sloan Foundation, the Participating Institutions, the National Science
Foundation, the U.S. Department of Energy, the National Aeronautics
and Space Administration, the Japanese Monbukagakusho, the Max Planck
Society, and the Higher Education Funding Council for England.The SDSS Web Site is http://www.sdss.org/.   

The SDSS is managed by the Astrophysical Research Consortium for the
Participating Institutions. The Participating Institutions are the
American Museum of Natural History, Astrophysical Institute Potsdam,
University of Basel, Cambridge University, Case Western Reserve
University, University of Chicago, Drexel University, Fermilab, the
Institute for Advanced Study, the Japan Participation Group, Johns
Hopkins University, the Joint Institute for Nuclear Astrophysics, the
Kavli Institute for Particle Astrophysics and Cosmology, the Korean
Scientist Group, the Chinese Academy of Sciences (LAMOST), Los Alamos
National Laboratory, the Max-Planck-Institute for Astronomy (MPIA),
the Max-Planck-Institute for Astrophysics (MPA), New Mexico State
University, Ohio State University, University of Pittsburgh,
University of Portsmouth, Princeton University, the United States
Naval Observatory, and the University of Washington.

\end{document}